\documentclass[aps,twocolumn,draft]{revtex4}
\input epsf
\begin{document}

\title{
Superconductivity without attraction in a quasi-one-dimensional metal
}
\author{A.V. Rozhkov}

\affiliation{
Institute for Theoretical and Applied Electrodynamics RAS,
Moscow, ul. Izhorskaya 13/19, 125412, Russian Federation
}


\begin{abstract}
An array of one-dimensional conductors coupled by transverse hopping and
interaction is studied with the help of a variational wave function. This
wave function is devised as to account for one-dimensional correlation
effects. We show that under broad conditions our system possesses the
superconducting ground state even if no attraction is present. The
superconducting mechanism is of many-body nature and deviates substantially
from BCS. The phase diagram of the model is mapped. It consists of two
ordered phases competing against each other: density wave, spin or charge, 
and unconventional superconductivity. These phases are separated by the
first order transition. The symmetry of the superconducting order parameter
is a non-universal property. It depends on particulars of the Hamiltonian.
Within the framework of our model possible choices are the triplet $f$-wave
and the singlet $d_{xy}$-wave. Organic quasi-one-dimensional superconductors
have similar phase diagram.
\end{abstract}
\date{\today}
\maketitle
\hfill

\section{Introduction}
In this paper we will study a system of one-dimensional (1D) conductors
arranged in a square lattice and coupled weakly in the transverse direction.
The purpose of this work is to show that in a rather general situation such
quasi-one-dimensional (Q1D) electron liquid with purely repulsive
electron-electron interaction is either a superconductor, or an insulator
with spin or charge density order. This is demonstrated with the help of a
certain variational wave function which adequately captures 1D many-body
effects.

The major issue in the description of the Q1D metal is the phenomenon of
dimensional crossover. At high energy the system can be viewed as a
collection of the Tomonaga-Luttinger (TL) liquids. However the TL liquid
cannot support a physical electron as an elementary excitation. Thus, at low
energy, where transverse single-electron hopping becomes important, it is
necessary to abandon the TL notions and use the Fermi-liquid approach
instead. Therefore, one is to stitch two different descriptions together to
obtain complete picture. 

From the technical point of view the source of the trouble is the conflict
between the many-body TL correlations and transverse single-electron
hopping, which is extremely difficult to handle within the framework of the
1D TL liquid \cite{boson}.

A simple method for the crossover description was proposed in 
Ref.\cite{rozhkov}. The latter method is based on a variational wave
function, whose generalized version we will use in this paper. 

To provide an intuitive introduction to the approach of \cite{rozhkov} we
briefly explain the structure of the variational wave function. Consider a
1D conductor described by the Tomonaga-Luttinger Hamiltonian. The ground
state of this system is the ground state of all TL bosons, with every
possible momenta $k_\|$ ($|k_\|| < \Lambda$, where $\Lambda$ is the cutoff of
the theory). Let's turn the transverse hopping on and
couple $N_\perp$ of these conductors into three-dimensional array. In this
situation the system will attempt to lower its ground state energy even
further by taking advantage of the transverse hopping energy. However, in
order to participate in hopping the bosons have to form many-body
fermion-like excitations, which have finite overlap with the physical fermion.

To accommodate for the possibility of having two types of excitations, bosonic
and fermionic, we device our variational state in the following fashion. We
introduce intermediate cutoff $\tilde\Lambda<\Lambda$. All TL bosons, whose
energy and momenta are high ($|k_\||>\tilde\Lambda$), remain in their ground 
states. The small momenta bosons ($|k_\||<\tilde\Lambda$) form fermion-like
excitations, which are delocalized in transverse direction. To distinguish
between the physical electrons and these fermionic excitations we will
refer to the latter as quasiparticles. In other words, the wave function
can be factorized into two parts. The high-energy part corresponds to the
ground state of $|k_\||>\tilde\Lambda$ Tomonaga-Luttinger bosons, the 
low-energy part corresponds to the three-dimensional anisotropic Fermi
liquid composed of the quasiparticles. 

The variational energy is minimized by adjusting $\tilde \Lambda$. The
energy of the quasiparticle transverse hopping is decreasing function of
$\tilde \Lambda$. At the same time, the in-chain energy grows when $\tilde
\Lambda$ grows. The trade-off between the transverse kinetic energy
and the in-chain potential energy determines the value of $\tilde \Lambda$.

If the optimal value of $\tilde \Lambda$ is non-zero, the low-energy
excitations of the system are the quasiparticles. Properties of the
fermionic quasiparticle state depend on quasiparticle effective Hamiltonian.
It arises naturally after high-energy bosons are `integrated out'. In this
effective Hamiltonian the anisotropy is insignificant. Due to this
circumstance, a standard mean field theory can be used to map out the
quasiparticle phase diagram.
Since the physical electron and the quasiparticle have finite overlap, there
is a direct correspondence between broken symmetry phases of the effective
Hamiltonian and the physical system. We already mentioned that the possible
phases of the system are spin-density wave (SDW), charge-density wave (CDW),
and superconductivity. The mean-field treatment of the effective
Hamiltonian was used in Ref. \cite{bour_caron} in order to demonstrate the
stability of the superconductivity in Q1D metals.

It is remarkable that the superconductivity is stable in the system with no
attraction. Superconductivity cannot be found if one apply mean-field
approximation to the bare Hamiltonian. To discover the existence of the
superconducting ground state the high-energy degrees of freedom must be
properly accounted for.

Our mechanism is related to that of Kohn and Luttinger. It is
known that classical Kohn-Luttinger mechanism gives extremely low
transition temperature. Our system does not share this property. Due to 
Q1D nature of the system the superconducting coupling constant is not as
minuscule as Kohn-Luttinger coupling constant. Consequently, the transition
temperature in our model does not have to be unobservable small.

We will see that the phase diagram of our model is similar to the phase
diagram of the organic superconductors: {\it (i)} when the nesting of the
Fermi surface is good, the ground state is either SDW or CDW; {\it (ii)}
under increased pressure the nesting is spoiled, the density wave becomes
unstable, and it is replaced by the unconventional superconductivity; 
{\it (iii)} under even higher pressure the superconducting transition
temperature vanishes, and the system shows no sign of the spontaneous
symmetry breaking. This similarity suggests that the proposed mechanism may
be relevant for these materials.

Yet, the purpose of this paper is not to model real-life systems. Indeed,
assumptions made about Hamiltonian's parameters may be too extreme for a real
material. Rather, we want to demonstrate in a controllable way that the
superconductivity in Q1D metals is a rather generic phenomena. Once this is
done, the qualitative understanding developed in a specialized model may be
applied to a more complicated situation, where analytical treatment is
problematic.

The paper is organized as follows. In Sect. \ref{model} we formulate our
model. In Sect. \ref{MF} we perform its mean-field analysis. The variational
calculations are done in Sect. \ref{var}. Different phases of the effective
Hamiltonian (and the physical system) are mapped in Sect. \ref{phase}. We
discuss the derived results in Sect. \ref{disc}.

\section{The system}
\label{model}

\subsection{The Hamiltonian }
We start our presentation by writing down the Hamiltonian for the array of
coupled 1D conductors:
\begin{eqnarray}
H&=&\int_0^L dx {\cal H},\label{H}\\
{\cal H}&=&\sum_{i} {\cal H}_i^{\rm 1D} + \sum_{i,j} 
\left[ {\cal H}_{ij}^{\rm hop} + {\cal H}_{ij}^{\rho\rho}
\right] ,
\end{eqnarray}
where the indices $i,j$ run over 1D conductors.
In this paper we will adhere to the agreement of denoting the Hamiltonian
densities with the calligraphic letters (e.g., ${\cal H}$) and full
Hamiltonians with the italic letters (e.g., $H$).

In the above formula the Hamiltonian density ${\cal H}_i^{\rm 1D}$
contains the in-chain kinetic energy and interactions:
\begin{eqnarray} 
{\cal H}_i^{\rm 1D}
&=&
{\cal T}^{\rm 1D}_i 
\left[
                \psi^\dagger ,
                \psi
\right]
+ 
{\cal V}^{\rm 1D}_i 
\left[
                \psi^\dagger ,
                \psi
\right]
+ 
{\cal V}^{\rm 1D}_{{\rm bs},i}
\left[
                \psi^\dagger ,
                \psi
\right],
\label{H1D}
\\
{\cal T}_i^{\rm 1D}
&=&
- {i} v_{\rm F} \sum_{p\sigma}
p\/\/ \colon\!
        \psi^\dagger_{p\sigma i} 
        (
             \nabla
             \psi^{\vphantom{\dagger}}_{p \sigma i}
        ) 
  \colon,
\label{T}
\\
{\cal V}^{\rm 1D}_i
&=&
g_2 \sum_{\sigma \sigma'} 
\rho_{{\rm L}\sigma i} \rho_{{\rm R}\sigma' i} 
+ 
g_{4} \left( 
                 \rho_{{\rm L}\uparrow i} 
                 \rho_{{\rm L}\downarrow i} 
                 + 
                 \rho_{{\rm R}\uparrow i} 
                 \rho_{{\rm R}\downarrow i}  
        \right) ,
\label{V}
\\
{\cal V}^{\rm 1D}_{{\rm bs},i}
&=&
g_{\rm bs}
\rho_{2k_{\rm F} i}\rho_{-2k_{\rm F} i},
\label{BS}
\end{eqnarray}
where the chirality index $p$ is equal to $+1$ ($p=-1$) for right-moving
(left-moving) electrons. The subscript `bs' stands for `backscattering'.
The theory has an ultraviolet cutoff $\Lambda = \pi/a$. The symbol 
$\colon \ldots \colon$ 
denotes the normal order of the fermionic fields with respect to the
non-interacting ground state. The Hamiltonian density 
${\cal H}^{\rm 1D}_i$ 
is spin-rotationally invariant.

Different densities used in formulae above and throughout the paper are
defined by the equations:
\begin{eqnarray} 
\rho_{p \sigma i}
&=&
\colon 
        \psi^\dagger_{p \sigma i}
        \psi^{\vphantom{\dagger}}_{p \sigma i} 
\colon,
\\
\rho_i
&=&
\sum_{p\sigma} \rho_{p\sigma i}, 
\\
\rho_{2k_{\rm F} i} 
& = & 
\sum_\sigma
\psi^\dagger_{{\rm R} \sigma i}
\psi^{\vphantom{\dagger}}_{{\rm L} \sigma i} ,
\\
\rho_{-2k_{\rm F} i}^{\vphantom{\dagger}}
&=&
\rho_{2k_{\rm F} i}^\dagger,
\\
{\bf S}_{2k_{\rm F} i}
&=&
\sum_{\sigma \sigma'} \vec{\tau}_{\sigma \sigma'} 
\psi^{\dagger}_{{\rm R} \sigma i}
\psi^{\vphantom{\dagger}}_{{\rm L} \sigma' i} ,
\\
{\bf S}_{-2k_{\rm F}i}^{\vphantom{\dagger}}
&=&
{\bf S}_{2k_{\rm F}i}^{\dagger}, 
\end{eqnarray}
where $\vec{\tau}$ is the vector composed of the three Pauli matrices.

The coupling between the 1D conductors is described by the transverse
terms: the single-electron hopping,
\begin{eqnarray}
{\cal H}_{ij}^{\rm hop}
&=&
- t(i-j) 
\sum_{p \sigma}
\left(
        \psi^\dagger_{p\sigma i}
        \psi^{\vphantom{\dagger}}_{p\sigma j} 
        +
        {\rm H.c.}
\right), 
\end{eqnarray} 
and the density-density interaction,
\begin{eqnarray} 
{\cal H}_{ij}^{\rho\rho}
&=&
g^\perp_0 (i-j) 
\rho_{i} \rho_{j} 
+ 
\label{perp}
\\
&&g_{2k_{\rm F}}^\perp (i-j) 
\left( 
          \rho_{2k_{\rm F} i}
          \rho_{-2k_{\rm F} j} 
          + {\rm H.c.}
\right).
\nonumber 
\end{eqnarray}
We accept that all interactions are repulsive, weak, and that the in-chain
interactions are stronger than the transverse interactions:
\begin{eqnarray}
2 \pi v_{\rm F}
\gg
g_{2,4}
\gg 
g_{\rm bs} 
\gg
g^\perp_0
\gtrsim
g^\perp_{2 k_{\rm F}}
> 0,
\label{hier}
\end{eqnarray}
and the transverse hopping is small:
\begin{eqnarray}
v_{\rm F} \Lambda \gg t.
\label{small_t}
\end{eqnarray} 
The constraints on the Hamiltonian's coefficients will be further discussed
in Sect. \ref{var}, Sect. \ref{superconductivity}, and Sect. \ref{accuracy}.

\subsection{Bosonized Hamiltonian}

In Sect. \ref{var} we will need the bosonized version of Hamiltonian
density ${\cal H}^{\rm 1D}$. The bosonic representation is based on the
bosonization prescription for the electron field \cite{boson}:
\begin{eqnarray}
\psi^\dagger_{p\sigma } (x)  
&=&
(2\pi a)^{-1/2} \eta_{p\sigma}
{\rm e}^{{\rm i}\sqrt{2\pi} \varphi_{p\sigma}(x)}, 
\label{bos}
\\
\varphi_{p\sigma} 
&=&
\frac{1}{2}
\left( 
	\Theta_c + p\Phi_c + \sigma \Theta_s + p\sigma \Phi_s 
\right).
\end{eqnarray}
In the above formulae 
$\eta_{p\sigma}$ 
is Klein factor, 
$\Theta_{c,s}$ 
are the TL charge ({\it c}) and spin ({\it s}) boson fields, 
$\Phi_{c,s}$ 
are the dual fields. The chain indices $i,j$ are omitted in the expressions
above. We will not show these indices explicitly in cases where such
omissions do not introduce problems. 

The bosonized one-chain Hamiltonian is:
\begin{eqnarray} 
{\cal H}^{\rm 1D} 
\left[
         \Theta,\Phi
\right]
= 
{\cal H}_0^{\rm 1D} 
\left[
         \Theta,\Phi
\right]
+
{\cal V}_{\rm bs}^{\rm 1D} 
[
	\Theta, \Phi
], \label{H1D_bos}
\end{eqnarray}
where ${\cal H}_0^{\rm 1D}$ is quadratic in the boson fields:
\begin{eqnarray}
{\cal H}_0^{\rm 1D} 
\left[
        \Theta,\Phi
\right] 
&=&
{\cal T}^{\rm 1D} 
\left[
        \Theta,\Phi
\right] 
+
{\cal V}^{\rm 1D}
\left[
        \Theta,\Phi
\right] 
\label{Hbos}
\\
\nonumber
&=&
\frac{v_c}{2} 
\left( 
        {\cal K}_c 
        \colon
               \left( 
                        \nabla \Theta_c 
               \right)^2
        \colon 
        + 
        {\cal K}^{-1}_c 
        \colon
               \left(
                        \nabla \Phi_c 
               \right)^2
        \colon 
\right)
\\ 
&&+
\frac{v_s}{2} 
\left( 
        \colon
               \left( 
                       \nabla \Theta_s 
               \right)^2
        \colon 
        + 
        \colon
               \left(
                       \nabla \Phi_s 
               \right)^2
        \colon 
\right), 
\nonumber 
\end{eqnarray}
while ${\cal V}^{\rm 1D}_{\rm bs}$ is not:
\begin{eqnarray} 
{\cal V}_{\rm bs}^{\rm 1D} 
[
    \Theta, \Phi
] 
&=&
\frac{g_{\rm bs}}{2\pi^2 a^2} 
\cos (
        \sqrt{8\pi} \Phi_s 
     )
\quad\\
&&-
\frac{g_{\rm bs}}{2\pi} 
\left[ 
        \colon
                \left(
                         \nabla \Phi_c 
                \right)^2
        \colon 
        +
        \colon
                \left(
                         \nabla \Phi_s 
                \right)^2
        \colon 
\right]. 
\nonumber 
\end{eqnarray} 
The symbol $\colon\ldots\colon$ denotes normal ordering of TL boson operators 
with respect to the non-interacting 
(${\cal K}_c = 1$, 
$v_{\rm s} = v_{\rm c}$, 
$g_{\rm bs} = 0$) 
bosonic ground state. The Tomonaga-Luttinger liquid
parameters are given by the formulae:
\begin{eqnarray}
{\cal K}_c 
&=&
\sqrt{
        \frac{2\pi v_{\rm F} + g_4 - 2g_2}
             {2\pi v_{\rm F} + g_4 + 2g_2}
},
\\
v_c 
&=& 
\frac{1}{2\pi} 
\sqrt{
         \left( 2\pi v_{\rm F} + g_4 \right)^2 - 4g_2^2
     },
\\
v_s 
&=&
v_{\rm F} - \frac{g_4}{2\pi}.
\end{eqnarray}
It is worth noting that 
\begin{eqnarray}
{\cal K}_c < 1,
\end{eqnarray} 
for repulsive interaction.

We will also need the expression:
\begin{eqnarray}
\psi^\dagger_{{\rm R} \sigma }
\psi^{\vphantom{\dagger}}_{{\rm L} \sigma }
=
\frac{1}{2 \pi a} 
{\rm e}^{
          {\rm i} 
          \sqrt{2\pi}
          (\Phi_c + \sigma \Phi_s)
        },
\end{eqnarray}
which gives operator
$\psi^\dagger_{{\rm R} \sigma }
\psi^{\vphantom{\dagger}}_{{\rm L} \sigma }$
in terms of the TL bosons.

\section{The mean-field approach}
\label{MF}

Once the model is formulated, it is not difficult to analyze its mean-field
phase diagram. Such analysis introduces serious qualitative errors. Yet, in
order to appreciate fully the advantage of the many-body calculations
proposed below the comparison with the mean-field results is very important.

From the outset we have to keep in mind that in our system several different
symmetries might be broken. Thus, several order parameters should be taken 
into consideration: SDW, CDW, triplet and singlet superconductivity. 

To perform the mean-field analysis we write the interaction terms as products
of these order parameters. After that the order corresponding to the
highest coupling constant and susceptibility is chosen.

\subsection{CDW and SDW}
\label{DW}

We start with the in-chain interaction (the biggest potential energy in the
system):
\begin{eqnarray}
\label{V_MF}
{\cal V}^{\rm 1D}_i
+
{\cal V}^{\rm 1D}_{{\rm bs},i} 
= 
-
\left(
        \frac{g_2}{2} - g_{\rm bs} 
\right) 
\rho_{{ 2k_{\rm F}}i}
\rho_{{ -2k_{\rm F}}i} 
\\
- 
\frac{g_2}{2} 
{\bf S}_{{2k_{\rm F}}i}
\cdot 
{\bf S}_{{ - 2k_{\rm F}}i} 
+ 
\ldots, 
\nonumber 
\end{eqnarray}
where $\ldots$ stand for $g_4$ term, which cannot be written as a product of
two order parameters. 

The transverse Hamiltonian may be expressed as a product of CDW and SDW
order parameters:
\begin{eqnarray}
\sum_{ij} 
{\cal H}^{\rho \rho}_{ij} 
\label{Hrr_mf}
&=&
\sum_{ij} 
g_{2k_{\rm F}}^\perp
\left( 
        \rho_{2k_{\rm F} i} 
        \rho_{-2k_{\rm F} j} 
        + {\rm H.c.}
\right)
\\
&-& g^\perp_0
\left( 
        \rho_{2k_{\rm F} ij}
        \rho_{-2k_{\rm F} ij}
        + 
        {\bf S}_{2k_{\rm F} ij}
        \cdot
        {\bf S}_{-2k_{\rm F} ij}
\right)  
+
\ldots.
\nonumber 
\end{eqnarray} 
The order parameter 
$\rho_{2k_{\rm F}ij}$ 
is equal to 
$\sum_\sigma
\psi^\dagger_{{\rm R} \sigma i}\psi^{\vphantom{\dagger}}_{{\rm L} \sigma j}$,
and 
${\bf S}_{2k_{\rm F}ij}$ 
is defined in a similar fashion. They are bond CDW and bond SDW. These types
of order cannot take advantage of the in-chain interaction energy (the
biggest interaction energy in the problem). Thus, they cannot compete against 
$\rho_{2k_{\rm F} i}$ 
and 
${\bf S}_{2k_{\rm F}i}$. 
We will not study 
$\rho_{2 k_{\rm F} ij}$ 
and 
${\bf S}_{2k_{\rm F}ij}$ 
anymore.

The non-interacting susceptibilities of SDW and CDW are equal to each
other.  Eqs. (\ref{V_MF}) and (\ref{hier}) suggest that the SDW coupling
constant is bigger than the CDW coupling constant:
\begin{eqnarray}
g_{\rm SDW} 
=
\frac{g_2}{2} 
>
g_{\rm CDW} 
&=&
\frac{g_2}{2} 
-
g_{\rm bs}
+
\frac{z^\perp g^\perp_{2k_{\rm F}}}{2},
\label{dw_coupling}
\end{eqnarray} 
where 
$z^\perp$
is the number of the nearest neighbours of a given chain. Thus, when the
nesting is good, the mean-field analysis suggests that the ground state is SDW.

\subsection{Superconducting orders}

Several sorts of the superconducting order parameter can be defined. 
They can be classified according to their spin and orbital symmetry.
It is useful to define a 
$2\times 2$ matrix 
$\hat \Delta_{ij}$ 
with components:
\begin{eqnarray}
\label{sc_matrix}
(\hat \Delta_{ij})_{\sigma\sigma'}
=
\psi^\dagger_{{\rm L}\sigma i}
\psi^\dagger_{{\rm R}\sigma' j},
\end{eqnarray}
and write
$\hat \Delta_{ij}$ 
as a sum of three symmetric matrices 
${ i}\vec{\tau}\tau^y$
and one antisymmetric matrix ${i} \tau^y$:
\begin{eqnarray}
\label{sc_order}
\hat \Delta_{ij}
= 
\frac{1}{\sqrt{2}}
\left[
           {\bf d}_{ij}
           \cdot 
           ({i}\vec{\tau}\tau^y) 
           +
           \Delta_{ij}
           {i} \tau^y
\right].
\end{eqnarray} 
The operator $\Delta_{ij}$ (${\bf d}_{ij}$) is the singlet (triplet) order
parameter corresponding to a Cooper pair composed of two electrons one of
which is on chain $i$ and the other is on chain $j$.

Furthermore, $\hat \Delta_{ij}$ may be symmetrized with respect to the chain
indices as well:
\begin{eqnarray}
\label{sc_symm}
\hat \Delta^{s/a}_{ij} 
=
\frac{1}{2}
\left(
          \hat \Delta_{ij} 
          \pm
          \hat \Delta_{ji} 
\right).
\end{eqnarray}
The superscript `s' (`a') stands for `symmetric' (`antisymmetric').

The operators $\Delta^{s/a}_{ij}$ and ${\bf d}^{s/a}_{ij}$ are defined in
the same fashion. If $i=j$, the antisymmetric quantities are, obviously, zero.
 
As the following derivations show, all these variants of superconductivity
are unstable at the mean-field level. The in-chain interaction energy can be
expressed as:
\begin{eqnarray}
{\cal V}_{i}^{{\rm 1D}}
+
{\cal V}_{{\rm bs},i}^{{\rm 1D}}
= 
(g_{2} - g_{\rm bs})
\mathbf{d}_{ii}^{\vphantom\dagger} 
\cdot
\mathbf{d}_{ii}^{\dagger} 
\\
+
(g_{2} + g_{\rm bs})
\Delta_{ii}^{\vphantom\dagger}
\Delta_{ii}^{\dagger} 
+ 
\ldots.
\nonumber 
\end{eqnarray} 
For realistic interaction $g_2 > g_{\rm bs}$. Therefore, the
one-chain order parameters ${\bf d}_{ii}$, $\Delta_{ii}$ are unstable.

The inter-chain interaction can be written as a bilinear of the
superconducting order parameters
${\bf d}_{ij}^{s/a}$, $\Delta_{ij}^{s/a}$, $i \ne j$:
\begin{eqnarray}
\label{sc_coupling}
&&\sum_{ij}    {\cal H}^{\rho\rho}_{ij}
=
\\
\nonumber 
&&\qquad
\sum_{ij} 
2 (g_0^\perp - g_{2 k_{\rm F}}^\perp) 
\left[ 
	\Delta_{ij}^{a}
	(\Delta_{ij}^a)^\dagger 
	+ 
	\mathbf{d}_{ij}^s
        \cdot
	(\mathbf{d}_{ij}^s)^{\dagger} 
\right] 
\\
&&\qquad
+ 
2 (g_0^\perp + g_{2 k_{\rm F}}^\perp)  
\left[ 
	\Delta_{ij}^{s}
	(\Delta_{ij}^s)^\dagger 
	+ 
	\mathbf{d}_{ij}^a
        \cdot
	(\mathbf{d}_{ij}^a)^{\dagger} 
\right] 
+ \ldots .
\nonumber 
\end{eqnarray}
For a realistic choice of the interaction constants: 
\begin{eqnarray}
g_{2k_{\rm F}}^\perp <  g_0^\perp. 
\label{bare_g}
\end{eqnarray} 
Consequently, the two-chain order parameters are unstable, as well as their
one-chain counterparts.

\subsection{Mean-field phase diagram}

As a result of the above considerations the following mean-field phase
diagram has emerged. If the nesting is good, the stable phase is SDW. It
is characterized by the non-zero $\langle {\bf S}_{2k_{\rm F}i} \rangle$.
The SDW state competes with the CDW state (non-zero
$\langle \rho_{2k_{\rm F}i} \rangle$).
SDW wins for it does not frustrate the backscattering interactions while
CDW does [see Eq.(\ref{V_MF})]. 

In a system with poor nesting SDW becomes
unstable \cite{sdw}. The mean-field theory predicts that such systems have
no spontaneously broken symmetry.

This phase diagram will be corrected in a qualitative manner when the
cooperative effects are accounted for. We will show that the many-body
phenomena force the violation of Eq.(\ref{bare_g}), which makes the
superconductivity stable in the systems with poor nesting. The same
phenomena may lead to violation of inequality (\ref{dw_coupling}), inducing
transition into CDW rather than SDW.

\section{Variational procedure}
\label{var}

In this section we develop the variational approach overcoming the
deficiencies of the mean-field approximation.

To keep our discussion short, transparent, and intuitive we will assume that
both backscattering and transverse interactions are zero: 
$g_{\rm bs} = 0$, $g^\perp_{0,{2k_{\rm F}}} = 0$. 
In such a situation the Hamiltonian is equal to:
\begin{eqnarray}
{H}' = \sum_i {H}_{0i}^{\rm 1D} + \sum_{ij} {H}_{ij}^{\rm hop}. 
\label{H'}
\end{eqnarray}
The first part of $H'$, the one-chain Hamiltonian 
$H_{0i}^{\rm 1D}$, 
is quadratic in terms of the TL bosons. The second part of $H'$, the
transverse hopping $H^{\rm hop}$, is quadratic in terms of the physical
fermion fields. Because of this circumstance, the variational derivations for
$H'$
are simpler than for generic $H$. Yet, such derivations retain the most
important features of the general case. This makes $H'$ an ideal object of
initial investigation, which we extend later for the Hamiltonian with non-zero 
$g_{\rm bs}$ 
and 
$g^\perp_{0,{2 k_{\rm F}}}$.

Below the prime mark ($'$) will be used to distinguish between the most
general Hamiltonian ${H}$, Eq.(\ref{H}), and the special case ${H}'$,
Eq.(\ref{H'}). Likewise, the prime will decorate the objects associated with
${H}'$ (e.g., effective Hamiltonian 
${H}^{{\rm eff}\prime}$,
variational energy
$E^{V\prime}$).

We first explain the heuristic idea behind our variational wave function.
Let us think of our system in terms of the TL bosons. The
bosonized version of 
${\cal H}^{\rm 1D}_0$ 
is given by Eq. (\ref{H1D_bos}).
However, the ground state 
$\left| 0_{\rm 1D} \right>$ 
of 
$H^{\rm 1D}_0$
is not a good approximation to the ground state of $H'$ for the finite-order
perturbation theory in $t$ is not well-defined.

On the other hand, if we were to describe our system with the help of the
bare electron degrees of freedom $\psi$, $\psi^\dagger$ we will not have
problems to account for $H^{\rm hop}$. But, within the fermionic framework,
the in-chain interaction energy is extremely difficult to handle.

To resolve this conflict we introduce the parameter
$\tilde \Lambda < \Lambda$ 
and separate the total phase space of the model into two parts, the
low-energy part (the degrees of freedom whose energy is smaller than 
$v_{\rm F} \tilde \Lambda$) 
and the high-energy part (the degrees of freedom whose energy is higher than
$v_{\rm F} \tilde \Lambda$) \cite{rozhkov}. 
The high-energy part will be described in terms of the TL bosons, while the
low-energy part will be described with the help of fermionic
quasiparticles, which we will define below. The exact value of 
$\tilde \Lambda$ 
is to be found variationally, as a trade-off between the in-chain
interaction and the transverse hopping. 

The formal implementation of this approach goes as follows. First, the TL
boson fields are split into two components: fast (with large momentum
$k_\|$: $\Lambda>|k_\||  > \tilde\Lambda$) and slow (with small momentum
$k_\|$: $|k_\||  < \tilde\Lambda$). The fast (slow) component will be
marked by `$>$' (`$<$') superscript:
\begin{eqnarray}
\Theta_{c,s} (x)
&=&
\Theta_{c,s}^<(x)   +   \Theta_{c,s}^>(x)
\\
&=&
\sum_{|k_\|| < \tilde \Lambda} 
                        \Theta_{c,s,k_\|} 
                        {\rm e}^{{\rm i} k_\| x}
+
\sum_{\tilde \Lambda < |k_\|| < \Lambda} 
                        \Theta_{c,s,k_\|} 
                        {\rm e}^{{\rm i} k_\| x},
\nonumber 
\\
\Phi_{c,s} (x)
&=&
\Phi_{c,s}^<(x)   +   \Phi_{c,s}^>(x)
\\
&=&
\sum_{|k_\|| < \tilde \Lambda} 
                        \Phi_{c,s,k_\|} 
                        {\rm e}^{{\rm i} k_\| x}
+
\sum_{\tilde \Lambda < |k_\|| < \Lambda} 
                        \Phi_{c,s,k_\|} 
                        {\rm e}^{{\rm i} k_\| x}.
\nonumber
%
\end{eqnarray} 
This split of the bosonic degrees of freedom induces the split of
the in-chain Hamiltonian density
${\cal H}_0^{\rm 1D}$:
\begin{eqnarray}
{\cal H}^{\rm 1D}_0 
                     \left[
                             \Theta,\Phi
                     \right] 
=
{\cal H}^{\rm 1D}_0 
                     \left[
                             \Theta^<, \Phi^<
                     \right] 
+
{\cal H}^{\rm 1D}_0 
                     \left[
                             \Theta^>,\Phi^>
                     \right].
\end{eqnarray}
That is, the Hamiltonian $H_0^{\rm 1D}$, Eq.(\ref{Hbos}), cleanly separates
into two parts corresponding to fast and slow modes.

The quasiparticles $\Psi_{p\sigma}^\dagger(x)$ are defined with the
help of Eq.(\ref{bos}), in which $a$ is substituted by
$\tilde a = \pi/\tilde\Lambda$ and the slow fields $\Theta_{c,s}^<$,
$\Phi_{c,s}^<$ or $\varphi_{p\sigma}^<$ are placed instead of the bare
fields $\Theta_{c,s}$, $\Phi_{c,s}$ or $\varphi_{p\sigma}$: 
\begin{eqnarray}
\Psi^\dagger_{p\sigma} (x) = (2\pi \tilde a)^{-1/2} \eta_{p\sigma}
{\rm e}^{{\rm i}\sqrt{2\pi} \varphi_{p\sigma}^<(x)}. \label{bosqp}
\end{eqnarray}
Using the quasiparticle field $\Psi_{p\sigma}$ we refermionize 
${\cal H}_0^{\rm 1D} [ \Theta^<, \Phi^< ]$:
\begin{eqnarray}
{\cal H}_0^{\rm 1D} 
= 
{\cal H}_0^{\rm 1D}
\left[
               \Psi^\dagger,
               \Psi
\right] 
+ 
{\cal H}_0^{\rm 1D} 
\left[
               \Theta^>,
               \Phi^>
\right],
\label{H1Dmix}
\\
{\cal H}_0^{\rm 1D}
\left[
               \Psi^\dagger,
               \Psi
\right] 
= 
{\cal T}^{\rm 1D}
\left[
                \Psi^\dagger ,
                \Psi
\right]
+
{\cal V}^{\rm 1D}
\left[
                \Psi^\dagger ,
                \Psi
\right],
\end{eqnarray}
where 
${\cal T}^{\rm 1D} [\Psi^\dagger, \Psi ]$ 
and 
${\cal V}^{\rm 1D} [\Psi^\dagger, \Psi]$ 
are given by Eqs.  (\ref{T}) and (\ref{V}).

The mixed representation of ${\cal H}_0^{\rm 1D}$, Eq.(\ref{H1Dmix}), makes
no sense in pure 1D problems since
${\cal H}_0^{\rm 1D}
\left[
          \Psi^\dagger,
          \Psi
\right]$
corresponds to an interacting 1D system, whose ground state and excitations
have no simple representation in terms of $\Psi$'s. Indeed, our variational
calculations will show that, if $t=0$, then 
$\tilde \Lambda = 0$. 
That is, no room for the quasiparticles is left in 1D situation. However, if
$t \ne 0$,
the quasiparticles delocalize in the transverse directions and lower the
total energy of the system. In such a case 
$\tilde \Lambda$ 
does not have to be zero, as we will demonstrate.

The Hamiltonian density ${\cal H}^{\rm hop}$ can be easily expressed within
the framework of the mixed quasiparticle-fast boson representation. One
observes that the physical fermion is simply:
\begin{equation}
\psi_{p\sigma}^\dagger 
= 
\sqrt{\tilde a/a}\Psi_{p\sigma}^\dagger 
{\rm e}^{{\rm i}\sqrt{2\pi} \varphi_{p\sigma}^>},
\label{physical}
\end{equation}
and that the fermionic and bosonic parts in this definition commute with
each other. Therefore:
\begin{eqnarray}
{\cal H}_{ij}^{\rm hop} 
= - 
\frac{\tilde a}{a} 
t 
\sum_{p \sigma}
              \Psi^\dagger_{p\sigma i}
              \Psi^{\vphantom{\dagger}}_{p\sigma j}
              {\rm e}^
              {
                   {\rm i} \sqrt{2\pi} 
                   (
                          \varphi_{p\sigma i}^>-\varphi_{p\sigma j}^>
                   )
              } 
              + 
              {\rm H.c.} 
\label{Hhopmix}
\end{eqnarray}
Eqs. (\ref{H1Dmix}) and (\ref{Hhopmix}) determine the form of
the total Hamiltonian ${H}'$ in the mixed representation. Let us study
this Hamiltonian.

The eigenenergies of the fast bosons are determined mostly by 
${\cal H}_0^{\rm 1D} \left[\Theta^>,\Phi^>\right]$. 
These eigenenergies are bigger than
$\sim v_{\rm F} \tilde \Lambda$. 
Small hopping term is only a correction to this quantity. Thus, we simply
neglect contribution of 
${H}^{\rm hop}$ 
to the high-energy sector's properties and assume that all fast bosons are
in the ground state 
$\left| 0_> \right>$
of the quadratic Hamiltonian:
\begin{eqnarray}
H^>
= 
\sum_i
\int {\cal H}_{0i}^{\rm 1D}     \left[ 
					\Theta^>, \Phi^> 
               		    \right]
dx.
\label{Hfast}
\end{eqnarray} 
When describing the quasiparticle state, we cannot neglect 
${\cal H}^{\rm hop}$:
the quasiparticles are low-lying excitations, and their energy
may be arbitrary small. Thus, we construct our variational wave function as
a product:
\begin{eqnarray}
\left| {\rm var} \right> 
= 
\left| \{ \Psi \} \right> 
\left| 0_> \right>,
\end{eqnarray}
where $\left| \{ \Psi \} \right>$ is the unknown quasiparticle state.
The variational energy is given by:
\begin{eqnarray}
\label{EV'}
E^{\rm V \prime}  
= 
\left< {\rm var} \right| 
	H' 
\left| {\rm var} \right>
= 
\left< \{\Psi\} \right| 
	H^{\rm eff\prime}
\left| \{\Psi \} \right>.
\end{eqnarray}
This equation defines the effective quasiparticle Hamiltonian $H^{\rm
eff \prime}$ as a `partial average' over the fast degrees of freedom:
\begin{eqnarray}
H^{\rm eff \prime}
&=&
\left<    0_>    \right| 
                              H^\prime
\left|    0_>    \right> 
\label{Heff'}
\\
\nonumber
&=&
H^{\rm 1D}_0 \left[
                     \Psi^\dagger , 
                     \Psi
           \right]
+
\tilde H^{\rm hop}\left[
                           \Psi^\dagger ,
                           \Psi
                  \right]
+
\left<    0_>    \right| 
                              H^>
\left|    0_>    \right>, 
\end{eqnarray}
where the last term is the $c$-number corresponding to the fast boson
contribution to the variational energy, and the effective quasiparticle
hopping in Eq.(\ref{Heff'}) is defined by the formula:
\begin{eqnarray} 
\tilde {\cal H}_{ij}^{\rm hop} 
&=&
- \tilde t 
\sum_{p\sigma}
            \Psi^\dagger_{p\sigma i}
            \Psi^{\vphantom{\dagger}}_{p\sigma j}
+ 
{\rm H.c.} ,
\\
\tilde t 
&=& 
t \frac{\Lambda}{\tilde \Lambda} 
\langle 
         {\rm e}^{{\rm i} \sqrt{2\pi} \varphi_{p\sigma }^>} 
\rangle_>^2.
\label{t_tilde0}
\end{eqnarray} 
The symbol $\langle \ldots \rangle_>$ is the short-hand notation for
$\langle 0_>| \ldots |0_> \rangle$.
The fast bosons
introduce renormalization of the effective hopping of the quasiparticles. The
expectation value in Eq.(\ref{t_tilde0}) is:
\begin{eqnarray}
\langle 
          {\rm e}^{{\rm i} \sqrt{2\pi} \varphi_{p\sigma }^>} 
\rangle_> 
=
\left( 
          \frac{\tilde \Lambda}{\Lambda} 
\right)^{({\cal K}_{\rm c} + {\cal K}_{\rm c}^{-1} + 2)/8}.
\label{expectation}
\end{eqnarray}
To establish the above equality we must remember that $\left| 0_> \right>$
is the ground state of the quadratic Hamiltonian $H^>$. Thus:
\begin{eqnarray}
&&\langle 
	{\rm e}^{{\rm i} \sqrt{2\pi} \varphi_{p\sigma }^>} 
\rangle_> 
=
{\rm e}^{-\pi \langle (\varphi_{p\sigma }^>)^2 \rangle_> },
\\
&&
\langle (\varphi_{p\sigma }^>)^2 \rangle_> 
= 
\\
&&\qquad
\frac{1}{4}
\left[
	\langle (\Theta_{c}^>)^2 \rangle_> 
	+
	\langle (\Phi_{c}^>)^2 \rangle_> 
	+
	\langle (\Theta_{s}^>)^2 \rangle_> 
	+
	\langle (\Phi_{s}^>)^2 \rangle_> 
\right]
\nonumber\\
&&\qquad\qquad
=
\frac{1}{8\pi}
\left[
	{\cal K}_c^{-1} + {\cal K}_c^{\vphantom{-1}} + 2
\right] 
\ln \frac{\Lambda}{\tilde \Lambda}.
\nonumber 
\end{eqnarray}
Substituting Eq.(\ref{expectation}) into Eq.(\ref{t_tilde0}) one finds:
\begin{eqnarray}
\tilde t 
= 
t 
\left( 
\frac{\tilde \Lambda}{\Lambda} 
\right)^
{({\cal K}_{\rm c} + {\cal K}_{\rm c}^{-1} - 2 )/4}.
\label{t_tilde}
\end{eqnarray} 
Assume now that the quasiparticle state
$\left| \{ \Psi \} \right>$
is non-interacting fermion ground state. Then the variational energy may be
expressed as follows:
\begin{eqnarray}
E^{V\prime}/ L N_\perp = \varepsilon^{\rm 1D} + \varepsilon^{\rm F},
\label{EV'_exp}
\end{eqnarray}
where $L$ is the length of the sample along the 1D conductors, $N_\perp$
is the number of these conductors; the one-dimensional contribution
$\varepsilon^{\rm 1D}$
and the non-interacting fermion contribution
$\varepsilon^{\rm F}$
are equal to:
\begin{eqnarray} 
\varepsilon^{\rm 1D} 
=
\frac{v_{\rm c} \theta}{2\pi} \left(\tilde
\Lambda^2 - \Lambda^2 \right),
\\
\varepsilon^{\rm F} 
=
- \frac{4}{\pi v_{\rm F}}\sum_i [ \tilde t(i)]^2
= 
- \frac{4}{\pi v_{\rm F}} 
\left(
	\frac{\tilde\Lambda}{\Lambda} 
\right)^{2\theta} \sum_i [ t(i)]^2,
\label{epsilonF}
\\
\theta 
=
\frac{1}{4} \left(
		{\cal K}_{\rm c}+ {\cal K}_{\rm c}^{-1} - 2 
	    \right).
\end{eqnarray}
Our expression for the fermion energy neglects all corrections coming from
quasiparticle interaction and possible symmetry breaking since these are
small.

It is convenient to define the characteristic transverse hopping energy as:
\begin{eqnarray}
\bar t^2 = \sum_i [t(i)]^2,
\end{eqnarray}
and the dimensionless ratio:
\begin{eqnarray}
\zeta = \frac{\tilde \Lambda}{\Lambda} < 1.
\end{eqnarray} 
In terms of such quantities the variational energy is equal to:
\begin{eqnarray} 
E^{V\prime}/ L N_\perp = \frac{v_{\rm c} \theta}{2\pi} \Lambda^2
(\zeta^2 - 1) -
\frac{4}{\pi v_{\rm F}} \zeta^{2\theta} \bar t^2.
\end{eqnarray} 
Minimizing it with respect to $\zeta$ one finds that for small in-chain
interactions ($\theta < 1$):
\begin{eqnarray}
\zeta = \left( \frac{8 \bar t^2}{ v_{\rm c} v_{\rm F} \Lambda^2}
\right)^{1/(2-2\theta)}. \label{zeta}
\end{eqnarray}
We see that, if $t=0$, the variational value of
$\tilde \Lambda$
is zero. In other words, in pure 1D system the quasiparticles do not appear.

Another important result obtained from Eq.(\ref{zeta}) is:
\begin{eqnarray}
\tilde t \sim v_{\rm F} \tilde \Lambda. 
\label{mf_cond}
\end{eqnarray}
This means that the anisotropy coefficient of the effective Hamiltonian
is of order unity: 
$(\tilde t/ v_{\rm F} \tilde \Lambda) \sim 1$. 
Therefore, the
mean-field treatment is appropriate for $H^{\rm eff\prime}$. The latter
conclusion is crucial for it signifies the completion of our quest: the
microscopic Hamiltonian $H'$, Eq.(\ref{H'}), whose treatment is complicated
by the presence of the 1D many-body effects, is replaced by the effective
Hamiltonian 
$H^{\rm eff \prime}$,
Eq. (\ref{Heff'}), which can be studied with the help of the mundane
mean-field approximation. 

Finally, we must extend the derivation of the effective Hamiltonian to the
situation of non-zero backscattering and transverse interactions. As
with the case of ${H}'$, the effective Hamiltonian $H^{\rm eff}$ for the
generic Hamiltonian $H$ is defined by the equation 
$ H^{\rm eff} = \left< H \right>_> $.
It is straightforward to show that $H^{\rm eff}$ has the same form as $H$
but with certain renormalizations of the coupling constants:
\begin{eqnarray}
\tilde g_2 = g_2,
\quad
\tilde g_4 = g_4,
\label{g24_tilde}
\\
\tilde g_{\rm bs} = g_{\rm bs},
\quad
\tilde g_0^\perp = g_0^\perp, 
\label{g_tilde}
\\
\tilde t 
=
\zeta^\theta t,
\quad
\tilde g_{2k_{\rm F}}^\perp 
= 
\zeta^{{\cal K}_c - 1}     g_{2k_{\rm F}}^\perp.
\label{g_2kF_tilde}
\end{eqnarray}
The derivations of these expressions are similar to the derivation of
Eq.(\ref{t_tilde}). For example, to calculate 
$\tilde g_{2k_{\rm F}}^\perp$ we must write:
\begin{eqnarray}
&&
\langle
	g_{2k_{\rm F}}^\perp 
	\rho_{2k_{\rm F}i}      \rho_{-2k_{\rm F}j}
\rangle_>
\\
&&\qquad\qquad
=
g_{2k_{\rm F}}^\perp 
\left( 
	\frac{\Lambda}{\tilde \Lambda}
\right)^2
\sum_{\sigma \sigma'} 
\Psi^\dagger_{{\rm R} \sigma i}
\Psi^{\vphantom{\dagger}}_{{\rm L} \sigma i}
\Psi^\dagger_{{\rm L} \sigma' j}
\Psi^{\vphantom{\dagger}}_{{\rm R} \sigma' j}
\nonumber
\\
&&\qquad\qquad
\times
\langle
	{\rm e}^
        {
             i\sqrt{2\pi} 
             \left[
                        (\Phi_{ci}^> - \Phi_{cj}^>) 
                        + 
                        (\sigma \Phi_{si}^> - \sigma' \Phi_{sj}^>) 
             \right]
        }
\rangle_>
\nonumber
\\
&&\qquad\qquad
=
\tilde g_{2k_{\rm F}}^\perp 
\sum_{\sigma \sigma'} 
	\Psi^\dagger_{{\rm R} \sigma i}
	\Psi^{\vphantom{\dagger}}_{{\rm L} \sigma i}
	\Psi^\dagger_{{\rm L} \sigma' j}
	\Psi^{\vphantom{\dagger}}_{{\rm R} \sigma' j},
\nonumber
\end{eqnarray}
where the effective coupling constant $\tilde g_{2k_{\rm F}}^\perp $ is
given by the expression:
\begin{eqnarray}
\tilde g_{2k_{\rm F}}^\perp 
=
g_{2k_{\rm F}}^\perp 
\left( 
	\frac{\Lambda}{\tilde \Lambda}
\right)^2
\langle
	{\rm e}^{i\sqrt{2\pi}[ 
				(\Phi_{ci}^> - \Phi_{cj}^>) 
			        + (\sigma \Phi_{si}^> - \sigma' \Phi_{sj}^>)
			     ] }
\rangle_> .
\end{eqnarray}
From this formula Eq.(\ref{g_2kF_tilde}) for 
$\tilde g^\perp_{2k_{\rm F}}$
follows. 

We want our effective Hamiltonian to be in the weak-coupling regime: when
the coupling is weak, the kinetic energy of the quasiparticles dominates
over their interaction, which justifies Eq.(\ref{epsilonF}). Consequently,
we need to impose a restriction on magnitude of the effective coupling
constants. Thus, in addition to Eq.(\ref{hier}) we require:
\begin{eqnarray}
\tilde g^\perp_{2 k_{\rm F}} \ll 2 \pi \tilde v_{\rm F}.
\label{geff_small}
\end{eqnarray}
Since
$\tilde g_{2k_{\rm F}}^\perp =  
g_{2k_{\rm F}}^\perp \zeta^{{\cal K}_c - 1}$,
inequality (\ref{geff_small}) is equivalent to:
\begin{eqnarray}
\frac{ {\bar t}}    {v_{\rm F}  \Lambda}
\gg
\left( 
	\frac{
		g^\perp_{2k_{\rm F}}
	     }       
	     {
		 v_{\rm F} 
	     }
\right)^{(1 - \theta)/(1 - {\cal K}_c )}.
\label{lower_b}
\end{eqnarray}
This gives the lower bound on the transverse hopping. In Sect.
\ref{accuracy}
we will explain how this inequality should be modified in order to improve
the accuracy of our method.

Keeping the above considerations in mind, one writes the equation for the
effective Hamiltonian:
\begin{eqnarray}
H^{\rm eff} = H^{\rm 1D} + \tilde H^{\rm hop} + \tilde H^{\rho \rho},
\end{eqnarray} 
where the tildes above 
$\tilde H^{{\rm hop}}$ 
and
$\tilde H^{\rho\rho}$ 
signify that the coupling constants of these terms are renormalized
according to Eqs.
(\ref{g24_tilde}), (\ref{g_tilde}), and (\ref{g_2kF_tilde}). 
The variational energy $E^V$ and $\zeta$ are given by Eqs.(\ref{EV'_exp}) and
(\ref{zeta}). The relation Eq.(\ref{mf_cond}) holds true for Hamiltonian
$H^{\rm eff}$ 
implying the applicability of the mean-field approximation.

\begin{figure} [!t]
\centering
\leavevmode
\epsfxsize=8cm
\epsfbox {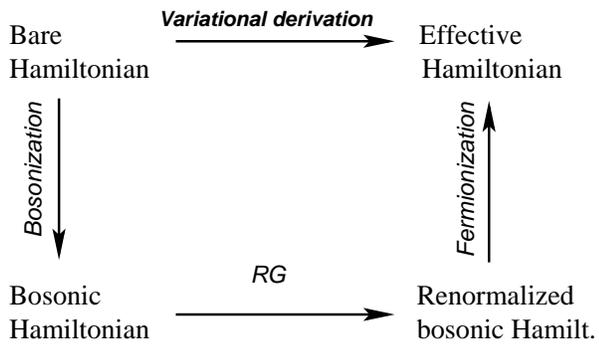}
\caption[]
{\label{diag} 
The relation between the variation procedure and the tree-level RG.
}
\end{figure}

Our variational derivation is equivalent to the tree-level renormalization
group (RG) result. Namely, one can execute the following sequence of
transformations. Beginning with the Hamiltonian $H$ one bosonizes it to
obtain the Tomonaga-Luttinger Hamiltonians for individual chains perturbed
by the transverse interactions, transverse hopping, and in-chain
backscattering. Because of the
presence of relevant (in RG sense) operators the RG flow takes the
Hamiltonian away from the Tomonaga-Luttinger fixed point. The flow must
be stopped when the renormalized transverse hopping amplitude becomes of
the order of the cutoff [see Eq.(\ref{mf_cond})]. At this point the bosonic
Hamiltonian has to be
refermionized. The resultant quasiparticle Hamiltonian coincides with
$H^{\rm eff}$. The relationship between the variational approach and the
RG procedure is shown on Fig.\ref{diag} in a form of a commutative
diagram.

This completes our derivation of the effective quasiparticle Hamiltonian
and we are prepared to analyze the phase diagram of our system.

\section{Phase diagram}
\label{phase}

How the phase diagram of the Hamiltonian $H$, Eq.(\ref{H}), can be
determined? It is essential to realize that the phase diagram of $H$
coincides with the phase diagram of $H^{\rm eff}$. Consider, for example,
the anomalous expectation value $\langle \psi^\dagger_{{\rm L} \uparrow i}
\psi^\dagger_{{\rm R} \uparrow j} \rangle$. For such a quantity the
following is correct:
\begin{eqnarray}
\langle 
	\psi^\dagger_{{\rm L} \uparrow i}
	\psi^\dagger_{{\rm R} \uparrow j} 
\rangle 
=
\left(
          \frac{\Lambda}{\tilde \Lambda}
\right)
\langle 
	\Psi^\dagger_{{\rm L} \uparrow i}
	\Psi^\dagger_{{\rm R} \uparrow j} 
\rangle
\langle
	{\rm e}^{i\sqrt{2\pi} \varphi_{p \sigma }^>}
\rangle_>^2.
\end{eqnarray} 
Since the bosonic expectation value is non-zero, both 
$\langle \psi^\dagger_{{\rm L} \uparrow i}
	\psi^\dagger_{{\rm R} \uparrow j} 
\rangle $ and 
$\langle \Psi^\dagger_{{\rm L} \uparrow i}
	\Psi^\dagger_{{\rm R} \uparrow j} 
\rangle $ are either simultaneously zero or simultaneously non-zero.
Same is true for other types of broken symmetries. This proves that the
phase diagram of $H$ and the phase diagram of $H^{\rm eff}$ are identical.
Since the properties of $H^{\rm eff}$ are accessible through the mean-field
approximation, we are fully equipped to explore the model's phase diagram.

\subsection{Density waves}

First, we consider the density wave phases. Both SDW and CDW have the same
susceptibilities but different effective coupling constants:
\begin{eqnarray}
\tilde g_{\rm SDW} = \frac{g_2}{2},
\\
\tilde g_{\rm CDW} = \frac{g_2}{2} - g_{\rm bs} + 
\frac{z^\perp \tilde g^\perp_{2k_{\rm F}}}{2}.
\end{eqnarray}
Due to strong renormalization of 
$\tilde g^\perp_{2k_{\rm F}}$,
inequality Eq. (\ref{dw_coupling}), which is always satisfied for bare
coupling constants, is not necessary fulfilled when the effective constants
are compared. Therefore, depending on the microscopic details, the density
wave phase could be of either nature. To be specific, we will study SDW
below. The discussion for CDW is completely the same.

SDW in Q1D metal was thoroughly analyzed at the mean-field level in Ref.
\cite{sdw}. We will follow this reference.

As we know, the stability of SDW depends crucially on the nesting of the
Fermi surface. Shape of the Fermi surface is determined by the effective
transverse hopping amplitudes 
$\tilde t (i)$.
If one assume that the only non-zero hopping amplitude
is the nearest neighbor amplitude $\tilde t_1$, then the resultant Fermi
surface nests perfectly. In order to describe the Fermi surface
with non-ideal nesting it is necessary to include at least the
next-to-nearest neighbor hopping amplitude $\tilde t_2$. For such structure
of hopping the SDW susceptibility is equal to:
\begin{equation}
\chi_{\rm SDW}   \approx
\frac{1}{\pi v_{\rm F}}
\times 
\cases{
         \ln\left( 2v_{\rm F}\tilde\Lambda/T \right),
	          & if $T>\tilde t_2=\zeta^\theta t_2$,\cr
         \ln\left( 2v_{\rm F}\tilde\Lambda/ \tilde t_2\right),
	          & if $T<\tilde t_2=\zeta^\theta t_2$.
      }
\label{chi}
\end{equation}
The SDW transition temperature is derived by equating 
$(g_2/2) \chi_{\rm SDW}$ 
and unity. For 
$\tilde t_2 = 0$
it is:
\begin{equation}
T_{\rm SDW}^{(0)} 
\propto 
v_{\rm F}\tilde\Lambda 
\exp\left(
		-2\pi v_{\rm F}/ g_2 
    \right).
\label{T0}
\end{equation}
If $\tilde t_{2} > 0$ the transition temperature $T_{\rm SDW}$ becomes
smaller then $T_{\rm SDW}^{(0)}$. It vanishes when
$\tilde t_{2} \propto T_{\rm SDW}^{(0)}$.
That is, exponentially small $\tilde t_2$ is enough to destroy SDW.

\subsection{Superconductivity}
\label{superconductivity}

The destruction of the density wave does not automatically implies that the
ground state becomes superconducting. By analogy with Eq.(\ref{sc_coupling})
we can write for the effective Hamiltonian:
\begin{eqnarray}
&&\sum_{ij}    \tilde {\cal H}_{ij}^{\rho\rho} 
=
\label{sc_eff}
\\
&&\quad
\sum_{ij} 
2
\left(
         \tilde g_0^\perp - \tilde g_{2k_{\rm F}}^\perp 
\right)
\left[ 
	\tilde \Delta_{ij}^{a}
	(\tilde \Delta_{ij}^{a})^\dagger 
	+ 
	\tilde \mathbf{d}_{ij}^{s} \cdot
	(\tilde \mathbf{d}_{ij}^{s})^{\dagger} 
\right] 
\nonumber 
\\
&&\qquad
+ 
2 (\tilde g_0^\perp + \tilde g_{2 k_{\rm F}}^\perp)  
\left[ 
	\tilde \Delta_{ij}^{s}
	(\tilde \Delta_{ij}^s)^\dagger 
	+ 
	\tilde \mathbf{d}_{ij}^a
        \cdot
	(\tilde \mathbf{d}_{ij}^a)^{\dagger} 
\right] 
+ \ldots,  \nonumber 
\end{eqnarray} 
where order parameters $\tilde \Delta_{ij}^{s/a}$ and ${\bf d}_{ij}^{s/a}$
are defined by Eqs.(\ref{sc_matrix}), (\ref{sc_order}), and (\ref{sc_symm}),
in which bare fermionic fields $\psi$ and $\psi^\dagger$ are replaced
by the quasiparticle fields $\Psi$ and $\Psi^\dagger$.

From Eq.(\ref{sc_eff}) we see that the effective superconducting coupling
constant 
$\tilde g_{\rm sc}$
is equal to:
\begin{eqnarray}
\tilde g_{\rm sc} 
=
2( \tilde g_{2k_{\rm F}}^\perp - \tilde g_0^\perp ).
\label{g_sc}
\end{eqnarray}
We conclude that the superconductivity is stable if:
\begin{eqnarray}
\tilde g_{2k_{\rm F}}^\perp >  \tilde g_0^\perp = g_0^\perp.
\label{tilde_g}
\end{eqnarray}
At the same time one has to remember that for the {\it bare} coupling
constants the inequality 
$g_{2k_{\rm F}}^\perp <  g_0^\perp$
holds true [see Eq.(\ref{bare_g})]. Can both inequalities be satisfied at the
same time? It is possible provided that the system is sufficiently
anisotropic. Indeed, the inequalities (\ref{tilde_g}) and (\ref{bare_g}) are
equivalent to:
\begin{eqnarray}
\frac{8 {\bar t}^2}    {v_c v_{\rm F}  \Lambda^2}
< 
\left( 
	\frac{g^\perp_{2k_{\rm F}}}          {g_0^\perp}
\right)^{(2-2\theta)/(1 - {\cal K}_c )} 
< 1.
\end{eqnarray}
This is the necessary condition for the superconducting ground state.
Similar condition was derived in \cite{rozhkov} for the spinless electrons.
This inequality gives an upper bound on $t$. This bound will be discussed
in Sect. \ref{accuracy} in connection with the method's dependability.

The final question is the type of the superconducting order realized
in our system. As one can see from Eq.(\ref{sc_eff}) there are two
candidates: singlet order parameter $\Delta_{ij}^a$ 
($d_{xy}$-wave
according to the accepted naming scheme \cite{review_RG1}) and triplet 
$\mathbf{d}^s_{ij}$
($f$-wave).
Both have the same coupling constant of 
$\tilde g_{\rm sc}$. 
This degeneracy cannot be lifted within our approach for we must include
subtler effects into our consideration. We will argue below 
(Sect. \ref{symmetry_subsection}) 
that the answer is sensitive to microscopic details of the system.
Therefore, in real materials either type of the superconductivity can be,
in principle, realized.

\subsection{Global phase diagram}

In this subsection we construct the global phase diagram of the system on the
pressure-temperature plane. 

The effect of the pressure on our Hamiltonian is twofold. First, it
increases the next-to-nearest neighbor hopping amplitude $t_2$. Thus, the
growth of the pressure spoils the nesting of the Fermi surface.

Second, it makes the system less anisotropic. This, in turn, leads to the
reduction of the 1D renormalization of 
$\tilde g^\perp_{2k_{\rm F}}$ 
under increasing pressure. Therefore, one can say that 
$\tilde g^\perp_{2k_{\rm F}}$ 
is decreasing functions of pressure.

Consequently, at low pressure the nesting is good and the ground state is
the density wave phase with the highest transition temperature possible: 
$T_{\rm SDW/CDW} = T^{(0)}_{\rm SDW/CDW}$. 
Under growing pressure the nesting property of the Fermi surface
deteriorates and 
$T_{\rm SDW/CDW}$
becomes smaller.

The density wave transition temperature decays until some critical pressure
$p_c$
at which it quickly goes to zero. At 
$p > p_c$ 
the subleading order, the superconductivity, is stabilized. The
characteristic superconducting critical temperature is smaller than
$T_{\rm SDW/CDW}^{(0)}$
for the density wave coupling constant is higher than that of the
superconductivity.  This is so because the density wave order benefits from
the in-chain interaction 
$g_2 \rho_{\rm L} \rho_{\rm R} $
while the superconductivity cannot do this.

The superconducting order parameter is either triplet ($f$-wave)
or singlet 
($d_{xy}$-wave). 
The superconducting gap vanishes at four nodal lines on the Fermi surface. 

Under even higher pressure $T_c \rightarrow 0$ for the system becomes less
anisotropic and inequality (\ref{tilde_g}) becomes invalid.

\begin{figure} [!b]
\centering
\leavevmode
\epsfxsize=8cm
\epsfbox {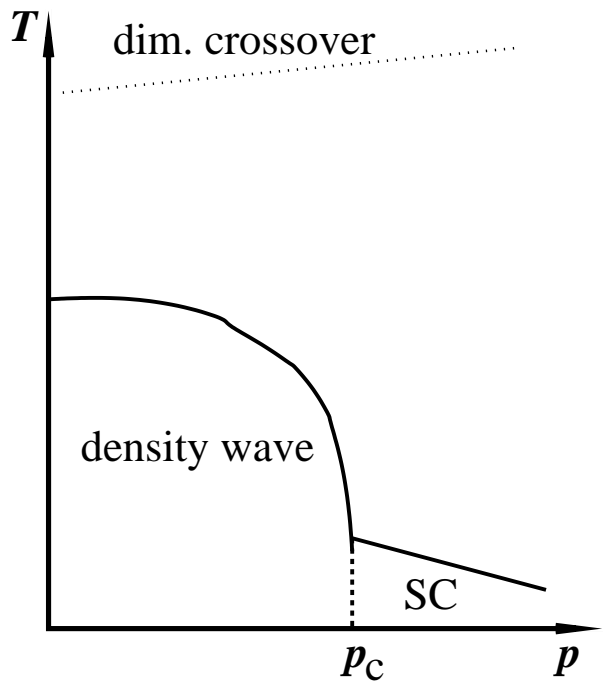}
\caption[]
{\label{fig1} 
Qualitative phase diagram of our model. Solid lines show second-order
phase transitions into density wave and the superconducting phases. Dashed
line shows the first-order transition between these phases. The dotted line
at high temperature shows location of the dimensional crossover.
}
\end{figure}

The schematic diagram is shown on Fig.\ref{fig1}.

\section{Discussion}
\label{disc}

This section is divided into four subsections. In subsection A we
discuss the accuracy of our method.

In subsection B we speculate under what condition 
$d_{x^2 - y^2}$-wave
superconductivity may be stabilized.

In subsection C we compare our approach with other theoretical methods
available in the literature.

In subsection D our theoretical results are compared against published
experimental data.

In subsection E we give our conclusions.

\subsection{Accuracy of the variational approach}
\label{accuracy}

In general, variational approach is an uncontrollable approximation, and
one may doubt our conclusions. Fortunately, the presented 
variational scheme is only a front for the tree-level RG transformation (see
Fig. \ref{diag}). Using RG notions, it is possible to prove rigorously that
the superconductivity is stable at least in a certain parameter range. Since
the stability of the superconducting phase depends on effective
inter-chain interactions, we must show that the tree-level RG is enough to
capture them adequately.

First, we must establish the structure of the tree-level RG flow. As implied
by Eq.(\ref{hier}), our model is near the Tomonaga-Luttinger fixed point 
($g_{\rm bs} = g^\perp_{0,2 k_{\rm F}} = 0$, $t=0$).
The fixed point Hamiltonian is perturbed by two relevant operators, $t$ and 
$g^\perp_{2 k_{\rm F}} $,
and one marginal,
$g_{\rm bs}$.
We assume that the transverse hopping is the most relevant operator: at the
dimensional crossover
($\tilde t \sim v_{\rm F} \tilde \Lambda$)
inequality (\ref{geff_small}) is satisfied.

This tree-level picture disregards several corrections. The most well-known
is
$(-g_{\rm bs}^2)$
contribution to the RG equation for 
$g_{\rm bs}$.
This term is of little immediate interest to us since it does not affect
the inter-chain interactions.

We identify three terms, which amend the RG equations for inter-chain
interaction constants. First, the backscattering contributes to the
anomalous dimension of 
$g^\perp_{2 k_{\rm F}}$:
the term proportional to
$g_{\rm bs} g^\perp_{2 k_{\rm F}}$
enters the RG equation for
$g^\perp_{2 k_{\rm F}}$.
This correction may be neglected since the anomalous dimension is
proportional to $g_2$, which is much larger than
$g_{\rm bs}$
[see (\ref{hier})].

Second, the in-chain interactions combined with the transverse hopping
contribute a term of order
$(g_{2,4}/v_{\rm F})^2 (t/v_{\rm F} \Lambda)^2$.
This term corrects inter-chain couplings by the amount:
\begin{eqnarray}
(\Delta g)_1 \sim 
\int_0^{\ell^*} d\ell 
\frac{
	[g_{2,4} t(\ell)]^2
     }
     {
	v_{\rm F}^3 \Lambda^2(\ell)
     }
\sim
\frac{
	(g_{2,4})^2
     }
     {
	v_{\rm F} 
     },
\\
t (\ell) = t {\rm e}^{- \theta \ell},
\end{eqnarray}
where $\ell$ denotes the scaling variable:
$\Lambda(\ell) = \Lambda {\rm e}^{-\ell}$.
The dimensional crossover occurs, and our our RG stops when $\ell$ reaches
the value
$\ell^* = \ln (\Lambda / \tilde \Lambda)$.
At the crossover it is true:
$t (\ell^*) / [ v_{\rm F} (\ell^*) \Lambda (\ell^*) ] 
= 
\tilde t / [v_{\rm F} \tilde \Lambda ] \sim 1$.

Third, the inter-chain interactions may contribute to the scaling equations
additional terms of order
$(g^\perp_{2 k_{\rm F}})^2$.
Such term corrects inter-chain coupling constants by the amount:
\begin{eqnarray}
(\Delta g)_2 \sim
\int_0^{\ell^*} d\ell 
\frac{
	[g^\perp_{2 k_{\rm F}} (\ell)]^2
     }
     {
	v_{\rm F} 
     }
\sim
\frac{
	(\tilde g^\perp_{2 k_{\rm F}})^2 
     }
     {
	g_{2}
     },
\\
g^\perp_{2 k_{\rm F}} (\ell) 
= 
g^\perp_{2 k_{\rm F}} {\rm e}^{ - ( 1 - {\cal K}_c )\ell },
\quad
1 - {\cal K}_c \sim g_2/v_{\rm F}.
\end{eqnarray}
Thus, the corrections to 
$\tilde g_{\rm sc}$
beyond the tree-level may be disregarded if
$\tilde g^\perp_{2 k_{\rm F}}$
is much bigger than
$(\Delta g)_{1,2}$.
This condition is equivalent to:
\begin{eqnarray}
(g_{2,4})^2 / v_{\rm F} 
\ll 
\tilde g^\perp_{2 k_{\rm F}} 
\ll 
g_{2}.
\label{tilde_g2}
\end{eqnarray}
We already derived inequalities binding 
$\tilde g^\perp_{2 k_{\rm F}}$
[see Eq.(\ref{geff_small}) and Eq.(\ref{tilde_g})]. Since 
$v_{\rm F}$
is bigger than
$g_2$,
Eq.(\ref{geff_small}) gives less restrictive upper bound on 
$\tilde g^\perp_{2 k_{\rm F}}$
than Eq.(\ref{tilde_g2}). Therefore, if we want an assurance that our method
does not lead us astray, we must
abolish Eq.(\ref{geff_small}) and use Eq.(\ref{tilde_g2}) instead.

The situation with Eq.(\ref{tilde_g}) is somewhat more complicated: it is
impossible to know, which quantity,
$g_0^\perp$
or
$(g_{2,4})^2 / v_{\rm F}$,
is smaller. Thus, we define:
\begin{eqnarray}
g_{\rm max} = \max \{
			(g_{2,4})^2 / v_{\rm F},
			g_0^\perp
		   \},
\label{g_max}
\end{eqnarray}
and rewrite Eq.(\ref{tilde_g2}) in the form:
\begin{eqnarray}
g_{\rm max} \ll \tilde g^\perp_{2 k_{\rm F}} \ll g_2.
\label{tilde_g3}
\end{eqnarray}
This inequality is self-consistent in the sense that
$g_{\rm max} \ll g_2$
[see Eq.(\ref{hier})]. It is convenient to cast Eq. (\ref{tilde_g3}) and
Eq.(\ref{small_t}) as a constraint on the bare hopping amplitude:
\begin{eqnarray}
\left(
	\frac{
		g^\perp_{2 k_{\rm F}}
	     }
	     {
		g_2
	     }
\right)^\frac{1 - \theta}{1 - {\cal K}_c}
\ll
\frac{ t }{ v_{\rm F} \Lambda }
\ll
\left(
	\frac{
		g^\perp_{2 k_{\rm F}}
	     }
	     {
		g_{\rm max}
	     }
\right)^\frac{1 - \theta}{1 - {\cal K}_c} \ll 1.
\label{t<>}
\end{eqnarray}
If this inequality is satisfied, then the model's phase diagram has a
superconducting phase, and the superconductivity is not an artifact of the
variational method. It is likely that some deviations from the constraints
imposed by Eqs.(\ref{hier}) and (\ref{t<>}) are not deadly for
superconductivity. Yet, they may affect the order parameter symmetry. This
issue is discussed in the next subsection.

\subsection{Symmetry of the superconducting order parameter}
\label{symmetry_subsection}

We have seen that the symmetry of the order parameter cannot be
unambiguously determined within the framework of our approximation:
as Eq.(\ref{sc_eff}) suggests, both $f$-wave and
$d_{xy}$-wave
states have similar energies. Our method capture only gross features of
the model, it is not delicate enough to calculate the superconducting
coupling constant with higher accuracy. We can identify at least two
mechanisms, which could lift the order parameter degeneracy. They work in
opposite direction. Thus, the final outcome depends crucially on the minutiae
of the microscopic model.

The mechanism promoting $f$-wave increases the coupling constant for this
order parameter and decreases the 
$d_{xy}$-wave
coupling constant. It operates in the following manner. The RG flow applied
to our system generates a new spin-dependent transverse interaction:
\begin{eqnarray} 
\tilde {\cal H}_{ij}^{SS}
=
\tilde J_{2k_{\rm F}}^\perp (i-j) 
\left( 
         \tilde {\bf S}_{2k_{\rm F} i} 
         \cdot
         \tilde {\bf S}_{-2k_{\rm F} j} 
         + 
         \text{H.c.} 
\right).
\end{eqnarray} 
At the dimensional crossover 
($\tilde t \sim \tilde v_{\rm F} \tilde \Lambda$)
one has:
$\tilde J_{{2 k_{\rm F}}} \sim g_{2,4}^2 / v_{\rm F}$.
This estimate can be found in, e.g., Ref. \cite{bour_caron}
[see first row, second column of Table I where 
$t'_\perp \sim E_0 (l)$].
The new term can be cast as:
\begin{eqnarray}
\label{sc_couplingSS}
\sum_{ij}    \tilde {\cal H}_{ij}^{SS} 
=
\sum_{ij}    
	     - 2\tilde J_{2k_{\rm F}}^\perp
\left[
	\tilde \mathbf{d}_{ij}^s
        \cdot
	(\tilde \mathbf{d}_{ij}^s)^{\dagger} 
        -3
	\tilde \Delta_{ij}^{a}
	(\tilde \Delta_{ij}^{a})^\dagger 
\right]
\\
+
2 \tilde J_{2k_{\rm F}}^\perp
\left[
	\tilde \mathbf{d}_{ij}^a
        \cdot
	(\tilde \mathbf{d}_{ij}^a)^{\dagger} 
        -3
	\tilde \Delta_{ij}^{s}
	(\tilde \Delta_{ij}^{s})^\dagger 
\right].
\nonumber 
\end{eqnarray} 
Thus, the $f$-wave coupling constant grows by 
$2 \tilde J^\perp_{2k_{\rm F}}$,
and the $d_{xy}$-wave coupling constant decreases by 
$6 \tilde J^\perp_{2k_{\rm F}}$.

A factor in favor of the $d_{xy}$-wave superconductivity is the
susceptibility. One can calculate two susceptibilities,
$\chi^{\rm sc}_f$
and
$\chi^{\rm sc}_d$,
for two order parameters:
\begin{eqnarray}
\chi^{\rm sc}_{f,d} 
=
\frac{1}{2 \pi v_{\rm F}}
\ln \left(
    		\frac{\tilde v_{\rm F} \tilde \Lambda}{T}
    \right)
+
C_{f,d},
\end{eqnarray}
where $C_{f,d}$ are the non-universal constants. In other words, the divergent
parts of both susceptibilities are identical, but the non-singular parts
depend on the order parameter symmetry and the band structure. 
This happens because our two orders have different orbital structure 
($f$-wave
is symmetric with respect to inversion of the transverse coordinate, while
$d_{xy}$-wave
is antisymmetric).

Within the
framework of our model (linear dispersion along the $x$-axis, square lattice,
small $\tilde t_2$) we have 
$C_f < C_d$. 
Thus, the susceptibility of $d$-wave is higher. 

The above analysis demonstrates that the symmetry of the order parameter is
a non-universal property very sensitive to the microscopic details.

It is reasonable to ask if one can stabilize
either of the remaining superconducting orders, ${\bf d}^a$ or $\Delta^s$, by
modification of the model's Hamiltonian. We can speculate that this might
be possible provided that the spin-spin interaction is enhanced.
Indeed, by examining 
Eq. (\ref{sc_eff}) 
and
Eq. (\ref{sc_couplingSS})
one concludes that 
$\Delta^s$
($d_{x^2-y^2}$-wave) 
could be non-zero if:
\begin{eqnarray}
3 \tilde J^\perp_{{2 k_{\rm F}}} > \tilde g^\perp_{{2 k_{\rm F}}}. 
\label{J>g}
\end{eqnarray} 
Such situation may be realized in a system with sufficiently large 
$g_{\rm bs}$ 
(to suppress CDW fluctuations) and sufficiently small
bare values of
$g_{2k_{\rm F}}^\perp$.

As for 
${\bf d}^a$,
it is always zero: the constants in front of 
${\bf d}^a$
are strictly positive in both Eq.(\ref{sc_eff}) and Eq.(\ref{sc_couplingSS}).

Thus, we demonstrate that the Q1D metal allows for a broad class of
superconducting orders. The choice between these orders depends on both the
band structure and the interaction constants.

\subsection{Other theoretical approaches}

The root of the superconductivity in the real-life Q1D materials
remains an unresolved
issue. It is often suggested that the superconductivity in these compounds
is not of phonon but rather of electron origin. There have been
many attempts to construct a mechanism in line with this suggestion.

The theoretical literature on the subject can be split into two groups
according to tools used. The studies employing the random phase approximation
(RPA) or the fluctuation exchange approximation (FLEX)
\cite{tanaka,kuroki,flex,review_RPA} constitute the first group. The second
group is made of the papers where RG
\cite{dup,nickel,nickelII,review_RG1,review_RG2}
is employed.

We have mentioned that our method is closely related to the RG
transformation. Clearly, it will be interesting to compare our conclusions
with the conclusions of other researchers who use similar strategies.

In Ref.\cite{dup,nickel,nickelII,review_RG1,review_RG2} the zero-temperature
phase diagram of
the Q1D metal was mapped with the help of a numerical implementation of the
one-loop RG flow. The authors of the latter papers were found that, if the
bare transverse interactions are zero or extremely small, the system
undergoes a transition from the SDW phase to the superconducting phase with
the order parameter
$\Delta^s_{ij}$ 
($d_{x^2-y^2}$-wave).

Furthermore, it was determined that, if the bare constants 
$g^\perp_{2k_{\rm F}}$ 
are sufficiently big, the transition is from the CDW phase into the
superconducting phase with the $f$-wave order parameter 
$\mathbf{d}^s_{ij}$.

The results of these papers can be understood within the framework of our
approach. In the limit where the only non-zero inter-chain term is the
transverse hopping ($t \ne 0$, \cite{dup}), the RG flow generates both 
$\tilde g^\perp_{2 k_{\rm F}}$
and 
$\tilde J^\perp_{2 k_{\rm F}}$.
These constants satisfy the relation Eq.(\ref{J>g}). The mechanism behind
this is described in the previous subsection.

As we pointed out, when Eq.(\ref{J>g}) is valid, the most
stable order parameter is 
$\Delta^s_{ij}$
($d_{x^2-y^2}$-wave).
Thus, our conclusions agrees with findings of 
Ref.\cite{dup}.

The limit studied in
\cite{nickel,nickelII}
is not compatible with our Eq.(\ref{hier}). In the latter reference it was
assumed that the in-chain backscattering is of the order of the in-chain
forward scattering. Thus, we cannot apply our approach straightforwardly,
but certain qualitative conclusions may be reached.

When bare 
$g^\perp_{2k_{\rm F}}$ 
is large, the effective coupling
$\tilde J^\perp_{2k_{\rm F}}$
is small, and the effective coupling 
$\tilde g^\perp_{2k_{\rm F}}$ 
is large. The ground state of the system with good nesting is CDW. The
destruction of the CDW phase takes place when the nesting becomes
sufficiently poor. Once the CDW is gone, we find ourselves in a familiar
situation where the stable superconducting order parameter is either 
$\mathbf{d}^s$ ($f$-wave) 
or 
$\Delta^a$
($d_{xy}$-wave), consistent with $f$-wave found in \cite{nickel,nickelII}.

If we lower 
$g^\perp_{2k_{\rm F}}$ 
sufficiently, the stability of SDW state may be restored
\cite{nickel,nickelII}. The in-chain backscattering suppresses 
$\tilde g_{2 k_{\rm F}}^\perp$
and promotes
$\tilde J^\perp_{2 k_{\rm F}}$, ultimately leading to inequality
(\ref{J>g}).
In such a regime the most stable order parameter is
$\Delta^s_{ij}$
($d_{x^2-y^2}$-wave), which agrees with \cite{nickel,nickelII}.

The above argumentation lends additional support to the notion that the
mechanism proposed in this paper is not an artifact of the variational
approximation.  It is also a convenient feature of our method that it is
analytical and the results of other approaches can be understood within its
framework.

Besides RG several authors use RPA or FLEX to determine the superconducting
properties in the anisotropic Fermi systems
\cite{tanaka,kuroki,flex,review_RPA}. These approximations resemble
the classical BCS scheme in which the phonons are
replaced by boson-like excitations of some other kind. In the quoted papers
the excitations mediating the attractive interaction between the electrons
are spin-density and charge-density fluctuations.

The frameworks laid out by the RPA and FLEX schemes are very appealing and
intuitive. They both predict that under certain condition the Q1D metal is
an unconventional superconductor. There is, however, a weak point: both
methods are unable to account for the peculiarities specific for 1D
electron liquid. Such weakness artificially narrows the region of the
parameter space where the superconductivity is stable.

Finally, the author recently developed a canonical transformation approach
for 1D electron systems 
\cite{rozhkovII,rozhkovIII,rozhkovIV}.
This method may be viewed as a generalization of the one discussed in this
paper. The application of the canonical transformation method to the Q1D
systems is in progress.

\subsection{Experiment vs. theory}

The question remains if the model and the mechanism discussed above are of
relevance to the Q1D superconductors, such as TMTSF and TMTTF
\cite{review_RG1}. Of cause, the latter compounds have very complicated
crystallographic structure: orthorhombic lattice, possibility of anion
ordering, dimerization \cite{book}. Yet, one can hope that these difficulties
are not of paramount importance as far as the superconducting mechanism is
concerned. 

If this hope is justified should be assessed by the mechanism's ability to
reproduce main features of the experimental data, at least qualitatively. We
can look at the presented model with a good degree of optimism for it
captures two most salient properties of the superconductivity in TMTSF/TMTTF.

The first of these two features is the common boundary shared by the
superconducting and the SDW phases on the pressure-temperature
phase diagram: the diagram of Fig.\ref{fig1} is similar to the
high-pressure part of the `universal' phase diagram of the TMTSF/TMTTF
compounds \cite{universal}.

The second is the non-trivial orbital structure of the order parameter in
the Q1D superconductors. There are numerous pieces of evidence in favor of
the order parameter with zeros on the Fermi surface
\cite{NMR,NMR2,imp,field,field2}. 
(However, there is a thermal transport measurement \cite{no_nodes} which
contradicts to this picture.) The order parameters 
$\mathbf{d}^s_{ij}$ 
and
$\Delta^{s,a}_{ij}$
are of this kind. Therefore, the predictions of our model is in qualitative
agreement with the experiment.

\subsection{Conclusions}

We proposed the superconducting mechanism for the strongly anisotropic
electron model without attractive interaction. We have shown that there is
a region in the parameter space where the superconductivity is stable and
shares a common boundary with SDW. The model supports two types of
unconventional superconducting order parameter. Our mechanism may be relevant
for the organic superconductors.

\section{Acknowledgements}

The author is grateful for the support provided by the Dynasty Foundation
and by the RFBR grants No. 06-02-16691 and 06-02-91200.


\end{document}